\documentclass[10pt,a4paper,reqno]{amsart}
\usepackage{amsthm,amssymb,amsmath,graphicx}
\usepackage{mathrsfs,setspace,pstricks,multicol,latexsym}
\usepackage{bm}
\usepackage{epsfig, subcaption, graphics}
\usepackage{hyperref}
\usepackage{multirow}
\usepackage{dsfont}
\usepackage{float}
\usepackage{rotating}
\usepackage{xfrac}
\advance\oddsidemargin by -1.6cm
\advance\evensidemargin by -1.6cm
\newtheorem{theorem}{{\bf Theorem}}
\newtheorem{lemma}[theorem]{{\bf Lemma}}

\newtheorem{defi}[theorem]{{\bf Definition}}
\newcommand{\noi}{\noindent}
\newcommand{\no}{\nonumber}
\begin{document}
\author{Anindya Goswami*}
\address{IISER Pune, India}
\email{anindya@iiserpune.ac.in}
\thanks{* Corresponding author}

\author{Kedar nath Mukherjee}
\address{National Institute of Bank Management Pune}
\email{kedar@nibmindia.org}

\author{Irvine Homi Patalwala}
\address{Department of Mathematics, University of Turin}
\email{zenvesta999@yahoo.com}

\author{Sanjay N. S.}
\address{IISER Pune}
\email{sanjayns100@gmail.com}

\title[Regime recovery in Markov modulated market model]{Regime recovery using implied volatility in Markov modulated market model}\thanks{This is supported by NBHM 02011/1/2019/NBHM(RP)R\& D-II/585, DST/INTDAAD/P-12/2020, DST FIST(SR/FST/MSI-105).}

\begin{abstract}
In the regime switching extension of Black-Scholes-Merton model of asset price dynamics, one assumes that the volatility coefficient evolves as a hidden pure jump process. Under the assumption of Markov regime switching, we have considered the locally risk minimizing price of European vanilla options. By pretending these prices or their noisy versions as traded prices, we have first computed the implied volatility (IV) of the underlying asset. Then by performing several numerical experiments we have investigated the dependence of IV on the time to maturity (TTM) and strike price of the vanilla options. We have observed a clear dependence that is at par with the empirically observed stylized facts. Furthermore, we have experimentally validated that IV time series, obtained from contracts with moneyness and TTM varying in particular narrow ranges, can recover the transition instances of the hidden Markov chain. Such regime recovery has also been proved in a theoretical setting. Moreover, the novel scheme for computing option price is shown to be stable.
\end{abstract}
\date{}		
\maketitle




\section{Introduction}
The regime switching extension is one of many generalizations of Black-Scholes-Merton model \cite{FBMS} of asset price dynamics. Regime switching models allow certain simplistic random variability of market parameters within finitely many possible states. Such models have been heavily studied in finance literature following the influential work of Hamilton \cite{hamilton}. In these, generally, the market parameters are modelled using Markov pure jump processes, whose states correspond to various Market regimes. Moreover, if the asset price evolves as a geometric Brownian motion (GBM) during the inter-transition period, such processes are called  Markov modulated geometric Brownian motion (MMGBM).
The fair-pricing of a range of derivatives including call and put options of European style in MMGBM or its extension have been extensively studied in the literature following \cite{Di}. For details, the readers may refer to  \cite{BGG}, \cite{JIPO}, \cite{ROTA}, \cite{FAN}, \cite{RR}, \cite{SIU}, \cite{SUX} and references therein. This list is merely indicative and not exhaustive. Some other works involving hidden Markov models of asset price, can be found in \cite{PEMA}, \cite{ROTA1}, and references therein.
Since the MMGBM market model is incomplete, the fair price of a derivative is not unique. In this paper, we would be dealing with the locally risk minimizing price (even if we omit the phrase ``locally risk minimizing'') which exists uniquely.

The fair price of a particular call option in MMGBM market depends on the current price of the underlying and the current market state. While the former is observed, the latter is not. Furthermore, the corresponding market parameter values and the state transition rates that influence the derivative price are also unobserved. However, in an efficient market, the option prices come with embedded information which may help in recovering the hidden variables and parameters, although not through an evident procedure. Needless to mention that historical data analysis offers an alternative route to infer the hidden market parameters. However, inferring the current state is not amenable to such analysis with only knowledge of stock price values. This is because, the unusual change in the stock price return could be due to a discontinuous change in the volatility coefficient or a rare occasion of a very large variation in Brownian motion term. Hence, it is essential to gather more information of asset price dynamics than the dynamics itself reflects. In this paper, we have addressed the problem of recovering the current regime using the derivative price data. As per our knowledge, the proposed approach is not present in the existing literature. A similar context with some interesting results and discussions in some different settings can be accessed through the papers of \cite{DEMO}, \cite{Jack}, and \cite{Ross}.

The main contribution of our paper is threefold. First, we have shown that the theoretical option price has sufficient information content for recovering the current regime. We have derived a precise formula for that. Second, by considering the realistic limitations of the unavailability of option price data in full spectrum, we have checked the applicability of the above theoretical result. This is a computational investigation and we have shown the promise of applicability. Thirdly, we have developed a novel numerical scheme for option pricing with good accuracy and efficiency. The stability analysis of the said scheme has also been added. In addition to these there are some other minor contributions related to producing numerical evidence of volatility smile effect under MMGBM market model. 

We have first obtained the implied volatility process of an MMGBM asset price process using the risk minimizing call option prices with fixed moneyness and  time to maturity (TTM). Then we show that the  implied volatility (IV) process is a pure jump process and is adapted to the filtration generated by the regime switching process. Furthermore, we have shown that if all regimes correspond to different option values, the regime switching process is also adapted to the IV process. We call this a recovery theorem due to its following implication. In a realistic situation the prices of call options of various strikes and maturities are known from the market data. So, the IV process is observable (as IV is computed using a deterministic function of observed quantities). Hence, although the regime process is unobserved, using the recovery theorem, the regimes are recoverable. The applicability of this theoretical result is debatable as not every day in a month is an expiration day of the call options, i.e., the obtainability of IV time series with fixed TTM for all consecutive days is unrealistic. One more implementation issue is the absence of options of all strike prices in the real market whereas stock price takes any values. So, IV process with fixed TTM and moneyness is not obtainable. Next we have relaxed the condition of constant TTM and moneyness and have allowed those to vary in an interval. Then we have experimentally validated that the perturbed IV process, thus obtained, can also recover the transition instances of the hidden Markov chain quite well. Our further numerical experiments have also revealed volatility smile. Ever since the 1987 crash of stock market, the volatility ``smile'' opens upward most of the time while the Black-Scholes model would require a flat volatility smile. By contrast, if an asset follows MMGBM, and its IV is computed based on the theoretical option prices, that is shown to exhibit a perfect ``smile'' in this paper. 

So far option pricing is concerned, it is known that the locally risk minimizing price function of any European vanilla option solves a partial differential equation (PDE) with an appropriate terminal condition. The domain of the PDE as well as the terminal data are unbounded for the Call option. Since the solution of the PDE has no known closed form expression \cite{MR}, one must truncate the domain before solving it numerically. This calls for truncation error which depends on the values of the artificially imposed boundary data. As per our knowledge, the relevant error analysis is absent in the existing literature. 
On the other hand, it is shown in \cite{GS} that the computational complexity for solving the PDE problem is significantly higher than that for solving an equivalent system of integral equations (IE) on a large bounded domain. Indeed, if $N$ and $M$ denote the time and space discretization sizes, the computational complexity of solving the PDE and IE are $O(NM^3)$ and $O(N^2M^2)$ respectively. So, we have derived and numerically solved an appropriate IE on a truncated domain using the asymptotic slope on the complement. As per our knowledge, this approximation of call option price function has not been studied in the literature. This equation produces a solution which is far more accurate or less expensive than those developed without considering the asymptotic slope. A stability analysis of the proposed scheme has also been included in this paper.

The rest of this paper is arranged in the following manner. In Section 2, we establish a measurability property of implied volatility for options with fixed moneyness and time to maturity. This result is crucial for regime recovery. For illustrating the implementability by numerical experiments, we first derive an integral equation on truncated domain for the price of European call option in Section 3.  Then we develop and study a numerical scheme for solving the integral equation on truncated domain in Section 4. A series of numerical experiments appear in Section 5, for illustrating various aspects of volatility smile, those emerge in the MMGBM market model. We show the possibility of regime recovery within the MMGBM framework using a family of numerical experiments in Section 6. Some concluding remark has been added in Section 7.  Some known statements, whose proofs are not readily available but have been used in the paper, are proved in the appendix.
\section{Implied volatility in the MMGBM Market Model}\label{sec1}
\subsection{Option Price in MMGBM Market Model} Let the drift and volatility parameters depend on the current state of the market, modelled as a Markov chain $X:=\{X_t\}_{t\geq 0}$ on the state space $\mathcal{X}:=\{1,2,\ldots, k\}$. We denote them as $\bm{\mu}=(\mu(1),\ldots, \mu(k))$ and $\bm{\sigma} =(\sigma(1), \ldots, \sigma(k))$ respectively. We further assume that $X$ is an irreducible Markov chain with a given rate matrix $\Lambda=(\lambda_{ij})_{k \times k}$. The one step transition probability to state $j$ given that the Markov chain left state $i$ is given by $p_{ij}: =\frac{\lambda_{ij}}{\lambda_{i}}$ for $i\neq j$, where $\lambda_i= |\lambda_{ii}|$. Thus $S_t$, the asset price at time $t$ satisfies
\begin{equation} \label{eq1}
   dS_t=\mu(X_t)S_t dt+\sigma(X_t)S_tdW_t\ \ \ \ \forall t>0, \ \ \ \ S_0>0.
\end{equation}
where $W:=\{W_t\}_{t\geq 0}$ is a standard Brownian motion on a given probability space $(\Omega, \mathcal{F},P)$. This process  $S:=\{S_t\}_{t\geq 0}$ is clearly a Markov modulated geometric Brownian motion (MMGBM). Here we keep the ideal bank's rate as a deterministic constant $r$. That means in this model, we consider regime switching in risky assets only. Although this market is arbitrage-free under admissible strategies, this is incomplete unlike the Black-Schole-Merton (BSM) model which assumes GBM dynamics of asset price \cite{Di}.

\noi Let $\varphi(t,s,i;K,T;r,\bm{\sigma}, \Lambda)$ be the (locally risk minimizing) price of European call option with strike price $K$ and maturity $T$ on a non-dividend paying underlying stock (satisfying \eqref{eq1}) at time $t$ when the stock price is $s$, and the market is at $i^{th}$ state. While $K$ and $T$ are the contract parameters, $r$, $\bm{\sigma}$, and $\Lambda$ are the model parameters. By suppressing the contract and model parameters, let us denote this function simply by $\varphi(t,s,i)$. It has been shown in \cite{Di}, and \cite{ADMKG} that the function $\varphi(t,s,i)$ satisfies the following system of parabolic partial differential equations (see for details),
 \begin{equation} \label{eq2}
  \frac{\partial \varphi(t,s,i)}{\partial t}+\frac{1}{2}\sigma(i)^2s^2\frac{\partial^2 \varphi(t,s,i)}{\partial s^2}+rs\frac{\partial \varphi(t,s,i)}{\partial s}+\sum_{j=1}^{k} \lambda_{ij}\varphi(t,s,j)=r\varphi(t,s,i)
\end{equation}
on $(0,T) \times (0,\infty) \times \mathcal{X}$ with the following terminal condition
\begin{align} \label{eq2a}
\varphi(T,s,i)=(s-K)^+, ~~ \forall s \in (0,\infty), i\in \mathcal{X}
\end{align}
where $\varphi(t,s,i)$ is of at most linear growth in $s$ and $x^+$ denotes $\max(0,x)$. If $\Lambda$ is a null matrix, meaning Markov chain $X$ does not transit almost surely, indeed the equation \eqref{eq2} coincides with that of classical BSM model. Some more results on \eqref{eq2}-\eqref{eq2a} are presented in Theorem \ref{theo1}. The paragraph below that mentions the relevant references.
\subsection{Implied Volatility Process}
\noindent For each $i \in \mathcal{X}$, let $C(t,s;K,T;r,\sigma(i))$, or $C_i(t,s)$ in short, denote the theoretical BSM price function of the option with strike price $K$, maturity $T$, fixed interest rate $r$, and fixed volatility $\sigma(i)$. To be more precise, BSM call option price is given by
\begin{equation}\label{BSMF}
  C( t,s;K,T;r,\sigma)=\Phi(d_1)s-Ke^{-r(T-t)}\Phi(d_2)
\end{equation}
where $d_1$ and $d_2$ are,
$$d_1=\left(\ln\left(\frac{s}{K}\right)+\left(r+\frac{\sigma^2}{2}\right)(T-t)\right)\bigg/\left(\sigma\sqrt{T-t}\right), \quad d_2=d_1-\sigma\sqrt{T-t}$$
and $\Phi$ is the CDF of the standard normal distribution \cite{FBMS}. A vital application of this price formula is found in defining the implied volatility of a risky asset. This fact is stated in the following theorem whose proof is added in the appendix for ready reference.
\begin{theorem}\label{ivtheorem}
Let $r$ be the constant risk-free interest rate of a friction-less, efficient market containing a stock. For every traded call option on that stock, the following equation
  \begin{equation}\label{IVdef}
      C( t,s;K,T;r,I)=M_p
  \end{equation}
can be solved for $I$ uniquely, where $C$ is as in \eqref{BSMF} and $K$, $T$ are the strike price and time of maturity of the option whose price is $M_p$ at time $t$ when stock price is $s$.
\end{theorem}
 We recall that the theoretical assumption on MMGBM model includes the frictionless and efficient market hypothesis. Since, in this market a fair price of the call option with strike price $K$ and maturity time $T$ is given by
$\varphi(t,s,i;K,T;r,\bm{\sigma}, \Lambda)$ at time $t$, when the stock price is $s$, Theorem \ref{ivtheorem} is directly applicable. Indeed, the implicit equation
\begin{align}\label{defIV}
C( t,s;K,T;r,I)= \varphi(t,s,i;K,T;r,\bm{\sigma}, \Lambda)
\end{align}
has a unique solution for $I$. The solution $I$ is the  implied volatility (IV) of the MMGBM asset price at time $t$, implied by a call option with strike price and maturity $K$ and $T$ respectively. In this section we construct an implied volatility process by selecting a class of call option contacts with fixed moneyness parameter. Subsequently we study the measurability properties of this process.
\begin{defi}\label{def5} Let $S=\{S_t\}_{t\ge 0}$ solve \eqref{eq1} and denote a stock price process. For any $p>0$, $\tau >0$, and $t>0$, let $I_t^{p,\tau}$ denote the IV of $S$ at time $t$ obtained using call option with strike $p$ times the present stock price $S_t$ and the TTM $\tau$. The process $I^{p,\tau}=\{I_t^{p,\tau}\}_{t\ge 0}$ is called the IV process with moneyness and TTM, $p$ and $\tau$ respectively. Here, $p$ less than, equal to, or greater than 1, correspond to in-the-money (ITM), at-the-money (ATM), or out-of-the-money (OTM) contracts respectively.
\end{defi}

\begin{theorem}\label{theo6}
Assume that the stock price process $S=\{S_t\}_{t\ge 0}$ follows MMGBM (Equation \eqref{eq1}). Then for every $p>0$, $\tau >0$, the implied volatility process $I^{p,\tau}$ (see definition \ref{def5}) is a pure jump process and is adapted to $X$ (filtration generated by $X$). If $\bm{\sigma}$, and $\Lambda$ are such that $i\mapsto \varphi(0,1,i,p,\tau, r, \bm{\sigma}, \Lambda)$ is an injection for some particular values of  $p>0$, and $\tau >0$, then $X$ is also adapted to $I^{p,\tau}$.
\end{theorem}

\proof Consider the regime switching GBM model as given in the section \ref{sec1}. We know that the locally risk minimizing price $\varphi(t,s,i;K,T;r,\bm{\sigma}, \Lambda)$ of a call option at time $t$ and $S_t=s$ with strike price $K$ and maturity time $T$ is given by
\begin{equation}\label{ivconstant1}
    \varphi(t,s,i;K,T;r,\bm{\sigma}, \Lambda)=E^*\Big(\frac{B_t}{B_T}(S_T-K)^+ \mid S_t=s, X_t=i\Big)
\end{equation}
where $B_t=e^{rt}$ and $E^*$ is the expectation with respect to the equivalent minimal martingale measure $P^*$ (see Lemma 2.1 \cite{BGG} for more details). Furthermore,
$S=\{S_u\}_{u\geq t}$ under $P^*$ solves
\begin{equation}
\nonumber dS_u=rS_u du+\sigma(X_u)S_udW_u^*  \ \ \ \, \forall u> t, \ \ \ \ S_t=s >0
\end{equation}
where $W^*$ is Brownian motion under $P^*$. Above equation has a closed form solution. We denote it by $S^{(t,s)}=\{S_u^{(t,s)}\}_{u\ge t}$. The solution is given by
\begin{equation}
\nonumber S_u^{(t,s)}=s \exp \Big(\int_{t}^{u}(r-\frac{\sigma(X_{t'})^2}{2})dt'+\int_{t}^{u}\sigma(X_{t'}) dW_{t'}^*\Big).
\end{equation}
Hence
$$S_u^{(t,1)}=\exp \Big(\int_{t}^{u}(r-\frac{\sigma(X_{t'})^2}{2})dt'+\int_{t}^{u}\sigma (X_{t'}) dW_{t'}^*\Big)$$
which implies,
$$S_u^{(t,s)}=sS_u^{(t,1)}.$$
Using the above equation, we rewrite equation \eqref{ivconstant1}.
\begin{align*}
   \varphi(t,s,i;K,T;r,\bm{\sigma}, \Lambda)=& E^*\Big(\frac{B_t}{B_T}(S_T^{(t,s)}-K)^+ \mid S_t^{(t,s)}=s, X_t=i\Big)\\
   =& E^*\Big(\frac{B_t}{B_T}s(S_T^{(t,1)}-\frac{K}{s})^+ \mid sS_t^{(t,1)}=s, X_t=i\Big) \\
   =& s\ E^*\Big(\frac{B_t}{B_T}(S_T^{(t,1)}-\frac{K}{s})^+ \mid S_t^{(t,1)}=1, X_t=i\Big)
   = s\ \varphi(t,1,i;\frac{K}{s},T;r,\bm{\sigma}, \Lambda).
\end{align*}
Since the BSM model is a special case of \eqref{eq1}, the BSM call option price $C( t,s;K,T;r,\sigma)$ also satisfies
$$C( t,s;K,T;r,\sigma)=s\ C( t,1;\frac{K}{s},T;r,\sigma).$$
Fix the moneyness parameter $p:=\frac{K}{s}>0$. From Definition \ref{def5}, implied volatility process $I_t^{p,\tau}$ solves the following equation
\begin{equation}\label{Iptau}
\varphi(t,S_t,X_t;pS_t,t+\tau ;r,\bm{\sigma}, \Lambda)=C( t,S_t;pS_t,t+\tau;r,I_t^{p,\tau}).
\end{equation}
By simplifying above using the preceding two equations, we get
    $$\varphi(t,1,X_t;p,t+\tau;r,\bm{\sigma}, \Lambda)=C( t,1;p,t+\tau;r,I_t^{p,\tau}).$$
Now using time homogeneity of the model we get
\begin{equation}\label{ivconstant2}
\varphi(0,1,X_t,p,\tau, r, \bm{\sigma}, \Lambda)= C(0,1,p,\tau, r, I_t^{p,\tau})
\end{equation}
for all time $t>0$. It is evident from \eqref{ivconstant2} that $I^{p,\tau}$ does not vary with time unless $X$ transits. More precisely, $I_t^{p,\tau}$ is adapted to $\mathcal{F}^X$ (filtration generated by $X$). The second assertion, i.e., $X$ is adapted to $I^{p,\tau}$, also follows from \eqref{ivconstant2} if $i\mapsto \varphi(0,1,i,p,\tau, r, \bm{\sigma}, \Lambda)$ is an injection. Hence the proof. \qed

\noindent While independence of $I_t^{p,\tau}$ with respect to the values of $S_t$ is not evident from the definition, that is clear from \eqref{ivconstant2}. This equation also suggests a numerical computation of $I^{p,\tau}$, when $X$ is observed. However, if $X$ is unobserved, and the prices of call options of all strike and maturity are known, one should consider \eqref{Iptau} for computing $I^{p,\tau}$.
\section{Computation of Option Price in the MMGBM Market Model}
\subsection{Integral Equation of Option Price}
The following theorem is taken from \cite{GS} and \cite{AJP}. This asserts that the MMGBM call option price function $\varphi(t,s,i)$ (as in \eqref{eq2}-\eqref{eq2a}) satisfies the integral equation \eqref{eq3}.
\begin{theorem} \label{theo1}
(i)The Cauchy problem \eqref{eq2}-\eqref{eq2a} has unique classical solution in the class of functions having at most linear growth. This is the locally risk minimizing price of call option with strike price $K$ and maturity $T$ at time t with $S_t=s, X_t=i$.\\
(ii) The following integral equation has a unique solution in the class of functions belonging to $C\Big([0,T] \times [0,\infty) \times \mathcal{X}\Big)$ $\cap$ $C^{1,2}\Big((0,T) \times (0,\infty) \times \mathcal{X}\Big)$ having at most linear growth
\begin{eqnarray} \label{eq3}
\varphi(t,s,i) = e^{-\lambda_i(T-t)} C_i(t,s)+\int_{0}^{T-t} \lambda_i e^{-(\lambda_i+r)v} \sum_{j\neq i} p_{ij}\int_{0}^{\infty}\varphi(t+v,x,j)\alpha(x;s,i,v)dx dv
\end{eqnarray}
for all $t \in [0,T]$, where for each $s>0$, $v>0$, $i\in \mathcal{X}$
$$x \mapsto \alpha(x;s,i,v) = \frac{\exp\left[-\frac{1}{2}\Big(\big(\ln(\frac{x}{s})-(r-\frac{\sigma(i)^2}{2})v\big)\frac{1}{\sqrt{v}\sigma(i)}\Big)^2\right]}{\sqrt{2\pi} \sigma(i) x\sqrt{v}}  $$
  is the pdf of Lognormal $\left(\ln s +\left(r-\frac{\sigma^2(i)}{2}\right)v, \sigma^2(i)v\right)$ distribution.\\
(iii) The solution $\varphi(t,s,i)$ of \eqref{eq3} solves \eqref{eq2}-\eqref{eq2a} classically.\\
(iv) For each $i\in \mathcal{X}$, $(s-K)^+\le \varphi(t,s,i) \le s$   for  all $t \in [0,T]$,  $s \in (0,\infty)$.
\end{theorem}
\noindent The proof of (i), (ii), and (iii) or a more general statement can be found in \cite{GS}, \cite{AJP} and \cite{DGP}. The proof of the second part of (i) can also be found in \cite{Di}, \cite{ADMKG}. For further discussion and proof of (iv) we refer to Proposition 4.1 in \cite{AJP}.
\noindent The integral equation \eqref{eq3} inherently includes all the boundary conditions. Because at $t=T$ the integral term on the right vanishes and hence
  $$\varphi(T,s,i)=C_i(T,s)=(s-K)^+.$$

\noindent To obtain the European call option price under MMGBM model one needs to solve either \eqref{eq2}-\eqref{eq2a} or \eqref{eq3}. Since it is not possible to solve these analytically,  we use numerical techniques to solve the integral equation.
\subsection{Asymptotic Behaviour of Call Price Function}
For solving numerically, the infinite integral in \eqref{eq3} has been replaced by a finite one in \cite{GS}. That approach relies on the fact that the infinite integral
$$\int_{M}^{\infty}\varphi(t+v,x,j)\alpha(x;s,i,v)dx$$
vanishes pointwise for an increasing value of $M$. However, to attain a certain degree of accuracy in numerical solution, $M$ needs to be considerably large,  which enlarges the size of truncated domain and enhances the time complexity at the quadratic order. Thus for  reducing the time complexity, we use the asymptotic behavior of the solution $\varphi$ in approximating the above mentioned infinite integral.
\begin{theorem} \label{ThAsympt}Let $\varphi$ solves \eqref{eq2}-\eqref{eq2a} then for each $t\in [0,T]$ and $i\in \mathcal{X}$
$$\lim_{s \to \infty}(s-\varphi(t,s,i))=Ke^{-r(T-t)}.$$
 \end{theorem}
\noindent The following lemma is useful for proving the above theorem.
\begin{lemma} \label{Lemma2} Let $\varphi$ be the solution of \eqref{eq3}. Then for all $(t,s,i) \in [0,T] \times [0,\infty) \times \mathcal{X}$,
  \begin{eqnarray}
\label{eq4}
s-\varphi(t,s,i) = e^{-\lambda_i(T-t)}(s-C_i(t,s))+\int\limits_{0}^{T-t} \lambda_i e^{-(\lambda_i+r)v}\sum_{j\neq i} p_{ij}\int\limits_{0}^{\infty}(x-\varphi(t+v,x,j))\alpha(x;s,i,v)dxdv.
\end{eqnarray}
\end{lemma}
 \proof We first note that  the mean of lognormal pdf $\alpha(\cdot;s,i,v)$ having the distribution\\
 $LN\bigg(\ln s + \Big( r - \frac{\sigma^2(i)} {2}\Big)v, \sigma^2(i)v\bigg)$, is equal to
\begin{equation}\label{eq5}
 \int_{0}^{\infty}x\alpha(x;s,i,v)dx =s e^{rv}.
\end{equation}
Hence,
\begin{align*}
\int\limits_{0}^{T-t} \lambda_i e^{-(\lambda_i+r)v}\sum_{j\neq i} p_{ij}\int\limits_{0}^{\infty}x\alpha(x;s,i,v)dxdv
  = \int\limits_{0}^{T-t} \lambda_i e^{-(\lambda_i+r)v} s e^{rv} dv
  =s\int\limits_{0}^{T-t} \lambda_i e^{-\lambda_iv}dv
  =s(1-e^{-\lambda_i(T-t)}).
  \end{align*}
  Now by adding both sides by $se^{-\lambda_i(T-t)}$, we get
  \begin{equation}\no
      s=s e^{-\lambda_i(T-t)}+\int\limits_{0}^{T-t} \lambda_i e^{-(\lambda_i+r)v}\times\sum_{j\neq i} p_{ij}\int\limits_{0}^{\infty}x\alpha(x;s,i,v)dxdv.
  \end{equation}
By subtracting the terms in \eqref{eq3} from those of the above equation, we get \eqref{eq4}.\qed

\noi A direct calculation involving the BSM European call option price formula gives the following limit expression
\begin{align}\label{BSa}
\lim_{s \to \infty}(s-C_i(t,s)) = Ke^{-r(T-t)} \ \ \ \ \forall i \in \mathcal{X}, \ \ t\in [0,T].
\end{align}

\noindent In other words,  for a fixed value of $K$ and $T$, the difference between stock price and the BSM option price converges to the present value of the exercise price. Next, we show that this is true for the MMGBM case.

\proof[Proof of Theorem \ref{ThAsympt}]
 Define $g_i(t) := \lim_{s \to \infty}(s-\varphi(t,s,i))$ for $t\in [0,T]$. So, from equation \eqref{eq4} and \eqref{BSa}, $g_i(t)$ is equal to  \begin{align*}
 &\lim_{s \to \infty}(s-\varphi(t,s,i))
     = K e^{-(\lambda_i+r)(T-t)} +\lim_{s \to \infty}\int\limits_{0}^{T-t} \lambda_i e^{-(\lambda_i+r)v} \sum_{j\neq i} p_{ij}\int\limits_{0}^{\infty}(x-\varphi(t+v,x,j))\alpha(x;s,i,v)dxdv.
  \end{align*}
  By using a substitution, $x=zs$, the right side of the above expression is equal to
  \begin{align*}
     & K e^{-(\lambda_i+r)(T-t)}+\lim_{s \to \infty}\int_{0}^{T-t} \lambda_i e^{-(\lambda_i+r)v} \sum_{j\neq i} p_{ij}\int_{0}^{\infty}(zs-\varphi(t+v,zs,j))\alpha(z;1,i,v)dzdv
  \end{align*}
where $\alpha$ is as in Theorem \ref{theo1}. Since, Theorem \ref{theo1}(iv) implies that $s-\varphi(t,s,i) \le K$ for all $s>0$, $i\in\mathcal{X}$, using dominated convergence theorem, we can pass the limit inside the integration and conclude that $g:=(g_1, \ldots, g_k)$ satisfies the following system of integral equations,
  \begin{equation} \label{eq6}
      g_i(t)=K e^{-(\lambda_i+r)(T-t)}+\int_{0}^{T-t} \lambda_i e^{-(\lambda_i+r)v}\times\sum_{j\neq i} p_{ij}g_j(t+v)dv
  \end{equation}
since the integration of $\alpha(\cdot;1,i,v)$ over $(0,\infty)$ is 1. Again as $r>0$, a direct application of Banach Fixed Point Theorem implies that \eqref{eq6} has  a unique solution for $g$. Then a direct substitution shows that $g_i(t)=Ke^{-r(T-t)}$ for all $i \in \mathcal{X}$ is indeed the unique solution of \eqref{eq6}. Thus
  $$\lim_{s \to \infty}(s-\varphi(t,s,i))=Ke^{-r(T-t)}.$$ \qed
\subsection{Option Price Equation on Truncated Domain}
\noi Now by splitting the domain of integration with respect to $x$ variable into two parts, equation \eqref{eq4} can be rewritten as
\begin{align*}
\varphi(t,s,i) = s-e^{-\lambda_i(T-t)}(s-C_i(t,s))-\int\limits_{0}^{T-t} \lambda_i e^{-(\lambda_i+r)v}\sum_{j\neq i} p_{ij}\Bigg(\int_{M}^{\infty}(x-\varphi(t+v,x,j))\alpha(x;s,i,v)dx \\
+ \int_{0}^{M} (x-\varphi(t+v,x,j)) \alpha(x;s,i,v)dx\Bigg)dv.
\end{align*}
Again, using the cdf of the standard normal distribution $\Phi$, we can write
\begin{equation}\label{eq:9} \int_{M}^{\infty}\alpha(x;s,i,v)dx = 1- \Phi\left(\frac{\ln{M/s}-(r-\frac{\sigma^2(i)}{2})v}{\sigma(i)\sqrt{v}}\right).
\end{equation}
For the sake of brevity we express the right side of \eqref{eq:9} as $1- F(M;s,i,v)$. Now by replacing the solution function $\varphi$, on interval $[M,\infty)$, by the asymptotic known function, we obtain an approximation $\varphi_M$ on the bounded domain $[0,T]\times [0,M]\times \mathcal{X}$ as the solution to the following integral equation
\begin{align*}
\varphi_M(t,s,i) =& s-e^{-\lambda_i(T-t)}(s-C_i(t,s)) - \int_{0}^{T-t} \lambda_ie^{-(\lambda_i +r)v}  Ke^{-r(T-t-v)} \left(1- F(M;s,i,v)\right) dv \\
& -\int_{0}^{T-t} \lambda_ie^{-(\lambda_i+r)v}\sum_{j\neq i} p_{ij}\int_{0}^{M}(x-\varphi_M(t+v,x,j)) \alpha(x;s,i,v)dx dv
\\
=& s-e^{-\lambda_i(T-t)}(s-C_i(t,s)) - Ke^{-r(T-t)}  \int_{0}^{T-t} \lambda_ie^{-\lambda_i v}  \left(1- F(M;s,i,v)\right) dv \\
& -\int_{0}^{T-t} \lambda_ie^{-(\lambda_i+r)v}\sum_{j\neq i} p_{ij}\int_{0}^{M}(x-\varphi_M(t+v,x,j)) \alpha(x;s,i,v)dx dv.
\end{align*}
It is clear that for any given $(t,s,i) \in [0,T] \times [0,\infty) \times \mathcal{X}$, if we allow $M\to \infty$, then the above equation converges to equation \eqref{eq4}, that is, $\varphi_M$ converges to $\varphi$ point-wise. However, the solution of the above equation need not be non-negative. Therefore we consider the following as the desired truncated problem
\begin{align} \label{eq10}
\no \varphi_M(t,s,i) =\max\Big( 0,
& s-e^{-\lambda_i(T-t)}(s-C_i(t,s)) - Ke^{-r(T-t)}  \int_{0}^{T-t} \lambda_ie^{-\lambda_i v}  \left(1- F(M;s,i,v)\right) dv \\
& -\int_{0}^{T-t} \lambda_ie^{-(\lambda_i+r)v}\sum_{j\neq i} p_{ij}\left(\int_{0}^{M}(x-\varphi_M(t+v,x,j)) \alpha(x;s,i,v)dx\right) dv \Big).
\end{align}
We consider the solution to equation \eqref{eq10} as an approximation of the price of the European call option. A numerical scheme is presented in the following section.
\section{Numerical Computation of Option Price}
\subsection{Numerical Scheme}
\noindent To solve equation \eqref{eq10} we use a step-by-step quadrature method. Let $\Delta_t$ and $\Delta_s$ denote the time step and stock price step size respectively so that $T=N\Delta_t$ and $M=M_0 \Delta_s$ for some $M_0$ and $N \in \mathbb{N}$. For every $n \in \{1,\ldots, N\}$, $l \in \{0,1,2,\ldots, N\}$, $m_0 \in \{1,2,\ldots, M_0\}$, $m \in \{1,2,\ldots, M_0\}$ we set
\begin{align*}
  G(m,m_0,l,i)=&\left\{\begin{array}{ll}
  \alpha(m_0\Delta_s,m\Delta_s,i,l\Delta_t)        & \textrm{ if } l\neq 0 \\
       \frac{1}{\Delta_s}& \textrm{ if } l= 0 \textrm{ and } m_0=m\\
       0& \textrm{ if } l= 0 \textrm{ and } m_0 \neq m,
  \end{array}\right.\\
\varphi^n_m(i)\approx&\varphi_M( T-n\Delta_t, m\Delta_s, i), \textrm{ and}\\
\varphi^n_0(i) = & 0.
\end{align*}
The last equality is due to part (iv) of Theorem \ref{theo1}. For a function $\Psi$ over the interval $[0,n\Delta_t]$, the quadrature rule is given by
\begin{equation} \no
    \int_{0}^{n\Delta_t}\Psi (v)dv= \Delta_t \sum_{l=0}^{n}w_n(l)\Psi(l\Delta_t)
\end{equation}
where for each $n\geq 1$, $w_n(1), \ldots, w_n(n)$ are the weights. The discretization of \eqref{eq10} using the above mentioned quadrature rule gives
\begin{equation} \label{eq11}
\varphi^n_m(i)=\max\left( 0, \xi^n_m(i) + \lambda_i \Delta_t w_n(0)\sum_{j\neq i} p_{ij}\varphi^n_m(j)\right)
\end{equation}
where
\begin{eqnarray*}
\xi^n_m(i):= m\Delta_s - e^{-\lambda_in\Delta_t}(m\Delta_s - C_i(T-n\Delta_t, m\Delta_s)) - Ke^{-rn \Delta_t} \lambda_i\Delta_t\sum_{l=1}^{n}w_n(l) e^{-\lambda_il\Delta_t} \\
\times (1-F(M_0 \Delta_s, m\Delta_s,i, l\Delta_t)) - \lambda_i\Delta_t w_n(0) m\Delta_s - \lambda_i \Delta_t \sum_{l=1}^{n}w_n(l)e^{-l(\lambda_i+r)\Delta_t} \\
\times \sum_{j\neq i} p_{ij}\Big(\sum_{m_0 =1}^{M_0}
\overline{w}(m_0)(m_0\Delta_s-\varphi^{n-l}_{m_0} (j))G(m, m_0,l,i)\Delta_s \Big)\\
\end{eqnarray*}
where $\overline{w}(1), \ldots, \overline{w}(M_0)$ are weights of Simpson's rule. We desire to fix $w_n$ according to the Simpson's rule also. However in Simpson's rule, interval of the integration must be divided into an even number of parts. While $M_0$ can be chosen to be even, $n$  can't be kept even always as $n$ runs from 0 to $N$ and takes odd values too. So we use a combination of Simpson's and trapezoidal rule for $w_n(\cdot)$.
To be more precise, for each $n\geq 1$, $w_n(l)=W(n,l+1)$ where the weight matrix $W$ looks like
$$W=
  \begin{bmatrix}
     \frac{1}{2}&  \frac{1}{2}\\
     \frac{1}{3} & \frac{4}{3}& \frac{1}{3} \\
     \frac{1}{3} & \frac{4}{3} & \frac{5}{6} & \frac{1}{2} \\
     \frac{1}{3} & \frac{4}{3} & \frac{2}{3} & \frac{4}{3} & \frac{1}{3} \\
     \frac{1}{3} & \frac{4}{3} & \frac{2}{3} & \frac{4}{3} & \frac{5}{6} & \frac{1}{2} &\cdots\\
       &  &  &  &  &  \vdots &\ddots
  \end{bmatrix}.$$
Here $n$ represents row number and $l$ represents column number. On $n(>1)$th row, for $n$ even, the first $n+1$ entries are wights of Simpson's rule and rest are zero; and for $n$ odd, the first  $n$ entries contain wights of Simpson's rule and $n$ and $n+1$ entries contain weights of trapezoidal rule.

\subsection{Stability Analysis}
\noindent As the first column of $W$ is nonzero, \eqref{eq11} is not an explicit equation. It is implicit as $\varphi^n_m(\cdot)$ appears on both sides of \eqref{eq11}. Moreover, the right side is not linear due to the presence of \emph{positive part} operator.
Let $\delta_n$ be the norm of single isolated perturbation in computing $\varphi^n$ and $\epsilon_q$ be its effect in computing $\varphi^{n+q}$. We seek for conditions on $\Delta_t$, for stability of the  numerical scheme under isolated perturbation. From the definition, naturally $ \epsilon_0=\delta_n $. For calculating $\epsilon_1$, we refer to equation \eqref{eq11}. As positive part is a non-expansion map, using triangular inequality we have
$\epsilon_1 \le (\epsilon' + \Delta_t\|\Lambda\| \epsilon_1)$, i.e., $\epsilon_1 \le \frac{\epsilon'}{1-\Delta_t\|\Lambda\|}$ where $\epsilon'$ is the effect of the error in $\xi^{n+1}$.
Since $\varphi^{n-l}$ is multiplied by lognormal density and  integrated, and also that all weights are less than 1,  we obtain $\epsilon'\le \epsilon_0 a\Delta_t$ where $a:= \max_i \lambda_i$. Thus, using $b:=\frac{a}{1-\Delta_t\|\Lambda\|}$ we get the following successive estimates
\begin{align*}
     \epsilon_1 \leq & \epsilon_0b\Delta_t\\
   \epsilon_2 \leq &(\epsilon_0+\epsilon_1)b\Delta_t \leq \epsilon_0b\Delta_t(1+b\Delta_t)\\
   \epsilon_3 \leq& (\epsilon_0+\epsilon_1+\epsilon_2)b\Delta_t \leq (\epsilon_0+\epsilon_0b\Delta_t+\epsilon_0b\Delta_t(1+b\Delta_t))b\Delta_t \leq \epsilon_0b\Delta_t(1+b\Delta_t)^2\\
   \vdots &\\
   \epsilon_q \leq &\epsilon_0b\Delta_t(1+b\Delta_t)^{q-1}.
\end{align*}
Let $\epsilon_T$ be total effect on $\varphi_m^N(i)$. then $\mid \epsilon_T \mid \leq \epsilon_{N-n}\leq \epsilon_0b\Delta_t(1+b\Delta_t)^{N-n-1}.$
As $\Delta_t=\frac{T}{N}$, if
\begin{equation}\label{stable}
b\Delta_t(1+\frac{bT}{N})^{N-n-1} \leq 1,
\end{equation}
then the effect would be less or equal to perturbation $\epsilon_0$ and the numerical scheme would be stable. On the other hand, if
\begin{equation}\label{error}
    \Delta_t \leq \frac{e^{-bT}}{b}
\end{equation}
then from the Lemma \ref{lemmaerror},  $\Delta_t \leq \frac{1}{b}\Big(1+\frac{bT}{N}\Big)^{-N}.$
Hence, for any $n \geq 0$, $b\Delta_t \leq \Big(1+\frac{bT}{N}\Big)^{-N+n+1}$ or, $b\Delta_t(1+\frac{bT}{N})^{N-n-1} \leq 1.$
This is the sufficient condition \eqref{stable} for the stability. Although right side of \eqref{error} also involves $\Delta_t$, that converges to a strict positive number as $\Delta_t$ tends to zero. Thus  \eqref{error} is a valid bound.
Let $\delta_n$ be bounded by constant $\delta$ for all $n$. Then total effect $\epsilon_T$ of the perturbation in the value $\varphi_m^N(i)$ is given by
$$\sum_{n=1}^{N-1}\epsilon_{N-n}<(e^{bT}-1)\delta.$$
Hence we obtain the following theorem.
\begin{theorem}
Let $b:=\frac{\max_i \lambda_i}{1-\Delta_t\|\Lambda\|} $. Under \eqref{error}, the numerical scheme is strictly stable with respect to an isolated perturbation. Moreover, the scheme displays uniformly bounded error propagation.
\end{theorem}
\subsection{Numerical Example}\label{numexp}
For the purpose of illustration, risk free interest rate $r$ is taken as $0.05$. This is close to the usual bank rate in real world. Rate matrix, drift parameter and volatility coefficients are
$$\Lambda=
  \begin{bmatrix}
    -10 & \frac{20}{3} & \frac{10}{3} \\
     10 & -20 & 10 \\
     \frac{10}{3} & \frac{20}{3} & -10
  \end{bmatrix},
\bm{\mu} =\begin{bmatrix}
       0.08 & 0.09 & 0.1
       \end{bmatrix} \textrm{ and } \bm{\sigma} = \begin{bmatrix}
         0.2 & 0.3 & 0.4
         \end{bmatrix}
         $$
respectively. We have restricted ourselves to the models where switching is not too fast and the volatility coefficient values at different regimes are not too close to each other. We set  strike price $K=1$ and the TTM $T$ as 25 days $\approx$ 0.1 year. Usually difference between successive maturity dates of monthly contracts of options are about 25 trading days. As, lesser the grid points, higher the computational error, to balance both error and time complexity, we choose $M=1.5$, $N=51$, and $M_0=400$. Here, $\Delta_t = \frac{T}{N}$ is indeed satisfying the stability condition \eqref{error}.


\begin{figure}[!ht]
	\centering
	\includegraphics[width=0.6\linewidth]{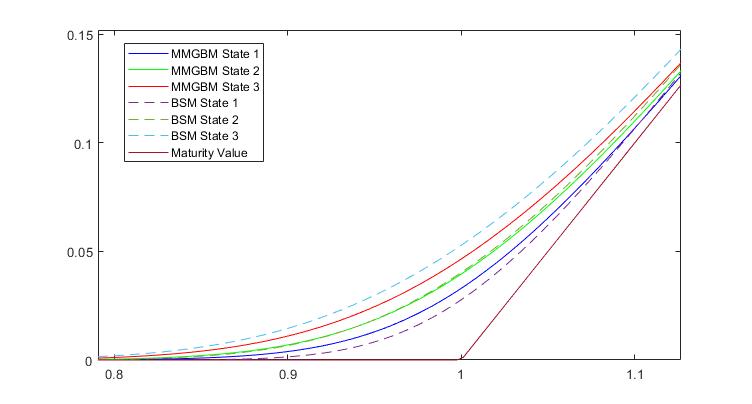}
	\caption{Call Option vs stock price with fixed strike and TTM}
	\label{optionprice}
	\centering
\end{figure}

\noindent Figure \ref{optionprice} exhibits the dependence of option price on stock price for a fixed contract. The stock and option price values are plotted along horizontal and vertical axes respectively. There are 7 line plots. Three solid and three dotted lines represent option prices at three different regimes under MMGBM and BSM models respectively. The lowest line represents maturity price of the option.
		
\section{Implied Volatility and Evidence of Smile in MMGBM Model}\label{volsmile1}
\subsection{Computation of Volatility Smile} We first show that MMGBM model can explain the stylized facts of volatility smile by considering the numerical example of Subsection \ref{numexp}. To this end we compute IV by solving \eqref{defIV}, where the right side is computed using the numerical scheme presented in Section 4.
\begin{figure}[h]
	\centering
	\includegraphics[width=10 cm]{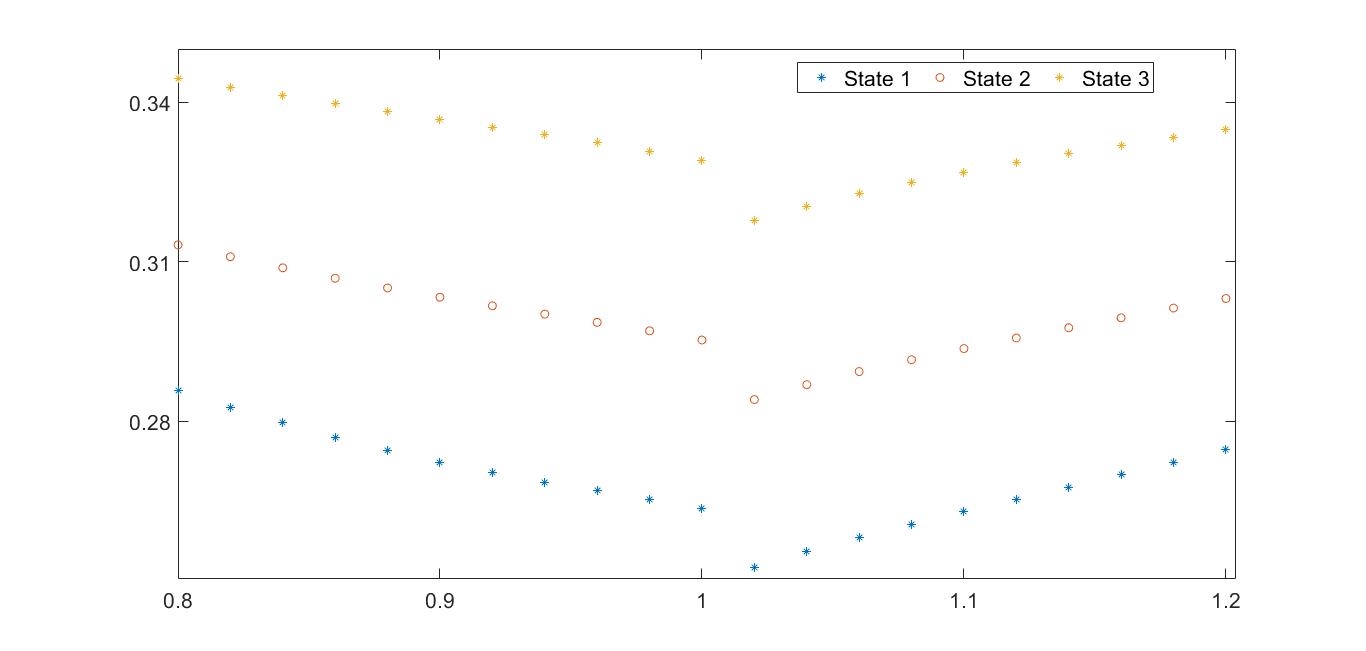}
	\caption{IV versus strike prices with fixed stock price}
	\label{smile}
	\centering
  \end{figure}
As prices of option depend on the regime value, the IVs depend too. So, IVs for all three regimes are computed for each strike price, ranging from 0.8 to 1.2 with an increment of 0.02 while fixing stock price as 1. Thus strike prices above or below 1 correspond to out of the money (OTM) call or put options respectively. In Figure \ref{smile}, the horizontal and the vertical axes represent strike price and the IV values respectively. By joining the IV values for each regime an upward curve can be obtained which is much similar to what is observed in the real option market.

\noindent Next, we attempt to numerically validate this observation for different sets of parameters.
For simplicity we consider regime-independent interest rate. We choose the extreme values of all parameters.
\noindent To be more precise, risk free interest rate $r \in \big\{0.01, 0.1\big\}$  (2 combinations), for all $i = 1, 2, 3$,  volatility coefficient $\sigma (i) \in \big\{0.1, 0.5\big\}$  (8 combinations), $\lambda(i) \in \big\{0.5, 3\big\}$  (8 combinations) and
  Transition probability matrix $$P = \begin{bmatrix}
    0 & \frac{2}{3} & \frac{1}{3} \\
     \frac{1}{2} & 0 & \frac{1}{2} \\
     \frac{1}{3} & \frac{2}{3} & 0
  \end{bmatrix}.$$
Out of these 128 possible combinations of parameters, in 32 cases all three sigma values are identical, which correspond to BSM cases. Those are not considered. For each of the remaining 96 cases, we vary strike prices (as required for Figure \ref{smile}) to obtain option prices at $s=1$, and the corresponding IV values. Now to identify the smile pattern, we use polynomial fit approach. Since the empirical volatility smile curve opens upward, we check positivity of the leading coefficient (we call that as smile coefficient) in the fitted quadratic regression equation.
  \begin{figure}[h]
	\centering
	\includegraphics[width=0.6\linewidth]{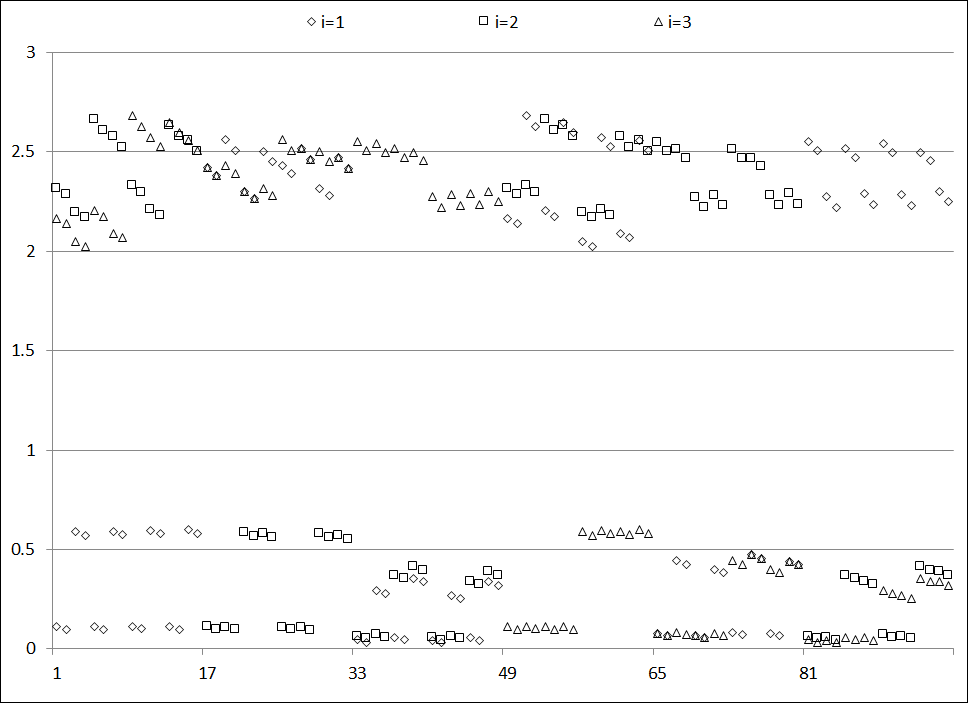}
	\caption{Smile Coefficient values for MMGBM asset  with different parameter combinations}
	\label{smilecoef}
	\centering
\end{figure}
In Figure \ref{smilecoef}, smile coefficients for all three regimes in all 96 cases are plotted and found to be positive. This validates that MMGBM model displays volatility smile for a wide range of realistic parameter values. In every consecutive pair, all parameters are kept identical except the $r$ values. Since those consecutive smile coefficient values are close to each other in Figure \ref{smilecoef},  we can infer that the shape of the smile curve is not much sensitive to interest rate $r$ values. We have further observed from the experimental outcome, in low volatility regimes, the smile coefficients are higher compared to high volatility regime. In other words, the smile curve tends to be flatter in high volatility regimes. Another observation is that the increase in lambda value increases the smile coefficient value.

\begin{figure}[h]
	\centering
	\includegraphics[width=0.6\linewidth]{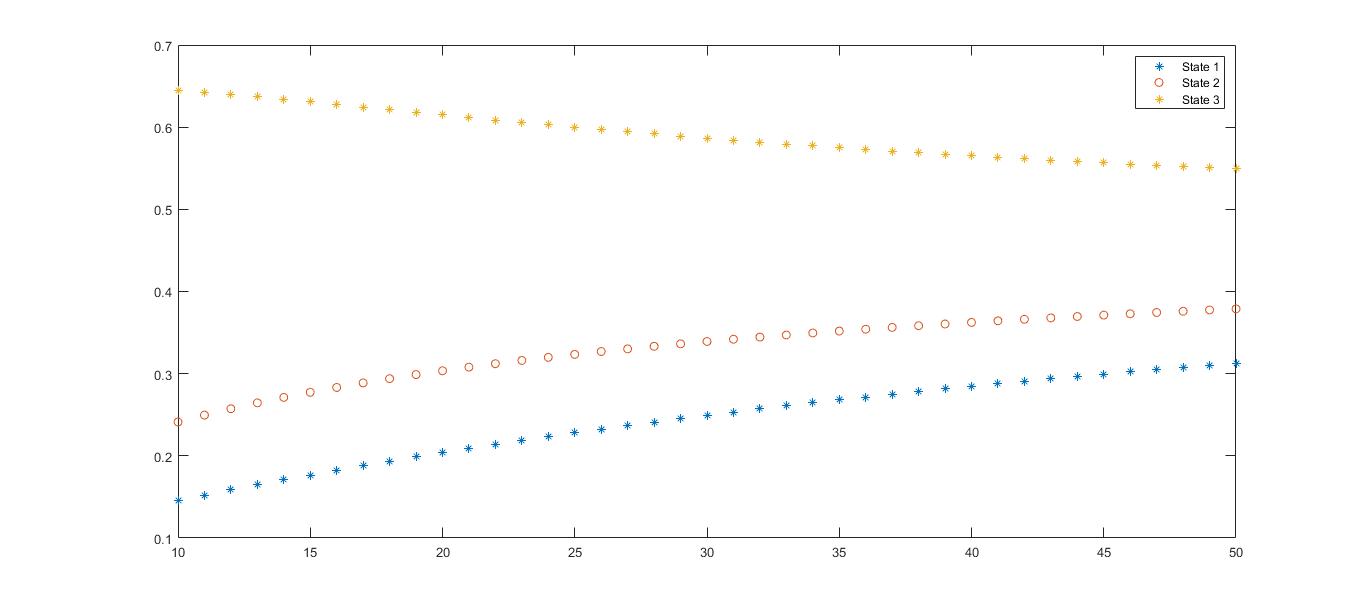}
	\caption{$ I^{1,\tau} $ vs $\tau$}
	\label{IVTTM}
	\centering
\end{figure}

\noindent Figure \ref{IVTTM} illustrates observation of another experiment. Here IV values for ATM options are plotted along the vertical axis against the TTM values ranging from 10 to 50 days along the horizontal axis.
We observe that IV decreases with TTM for the lowest volatility regime whereas that increases for highest volatility regime with decrease in TTM. The same pattern has been observed for in the money(ITM) and out of the money (OTM) options also.
\subsection{Computation of IV process and Quality of Approximation}
In a realistic situation, the regime process $X$ is unobserved. Nevertheless, the prices of call options of various strikes and maturities are known from the market data. Therefore, for finding $I^{p,\tau}$, one should instead of \eqref{ivconstant2}, consider \eqref{Iptau} where the left side value can be observed. However, for numerical investigations, that term should be generated numerically. As that has no closed form expression, we obtain $I_{t}^{S_t,pS_t,\tau}$, an approximation of $I_t^{p,\tau}$, satisfying
\begin{align*}
\varphi_M(t,S_t,X_t;pS_t,t+\tau ;r,\bm{\sigma}, \Lambda)=C( t,S_t;pS_t,t+\tau;r,I_t^{S_t,pS_t,\tau}).
\end{align*}
\begin{figure}[h]
	\centering
	\includegraphics[width=0.6\linewidth]{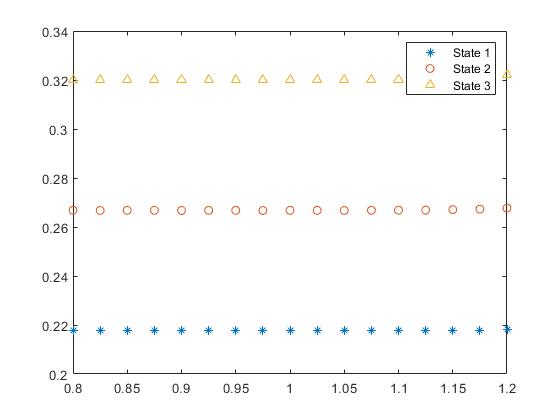}
	\caption{$\bar{I}_i(s)$ vs $s$}
	\centering
	\label{ATMIV1}
\end{figure}
We call $\{I_t^{S_t,pS_t,\tau}\}_{t\ge 0}$ as the approximated IV process or AIVP in short. It is known that small perturbation in option prices may result in large changes in IV, as the Vega in BSM model is pretty small. Hence, it is not evident if the constancy of $\bar{I}_i(s):= E[I_t^{S_t,pS_t,\tau}\mid S_t=s, X_t=i]$ in $s$ (as in Theorem \ref{theo6}) is likely to be observed for a limited precision in $\varphi_M$. Below we observe that the computational error $\|\varphi -\varphi_M\|$ in the example of Subsection \ref{numexp} is sufficiently small for successfully identifying constancy of $\bar{I}_i(\cdot)$. For a numerical illustration, we fix $p=1$, $\tau =0.1$ and vary stock price $s$ and regime $i$ from 0.8 to 1.2 with step size 0.025 and from 1 to 3 respectively. For each case, we evaluate $\bar{I}_i(s)$ which is plotted along the vertical axis in Figure \ref{ATMIV1}. This illustrates the anticipated constancy for that specific example.
Next, for examining the above mentioned constancy of $\bar{I}_i(\cdot)$, we set for each $i\in \mathcal{X}$, the relative error
$$\bar{e}_i:=\left( \max_s \bar{I}_i(s) - \min_s \bar{I}_i(s) \right)/\min_s\bar{I}_i(s)$$
and find its maximum over a wide variety of parameter combinations. To this end, we use the same set of parameter combinations that has been used for producing Figure \ref{smilecoef}. In order to estimate $\bar{e}_i$, corresponding to each combination of parameter values, $\bar{I}_i(s)$ is computed for each $s$ from 0.8 to 1.2 with step size 0.025, for all $i=1,2,3$. The maximum error, obtained for any regime and for any set of parameters is of the order of $10^{-4}$ which is nominal. Above numerical evidences indicate that the AIVP $\{I_t^{S_t,pS_t,\tau}\}_{t\ge 0}$ is a reasonable numerical approximation of $I^{p,\tau}$ process. We use AIVP in subsequent sections.
\section{Recovery of Transitions in MMGBM using AIVP}
\subsection{Algorithmic Aspect}
Following is the step by step procedure to obtain AIVP time series on a simulated MMGBM data of stock prices.
\begin{enumerate}
    \item For a given rate matrix we simulate a continuous time Markov chain $X$.
    \item Using the generated Markov chain, we simulate stock prices by discretizing the solution of \eqref{eq1}. To be more precise, we make use of
        \begin{equation}
            S_{t+h}=S_{t}\exp\left[\left(\mu(X_t) -\frac{\sigma(X_t)^2}{2}\right)h+\sigma(X_t)\sqrt{h}Z_t\right]
        \end{equation}
repeatedly, where $\mu$ and $\sigma$ correspond to drift and volatility parameters, $h$ is a positive time step,  $t\in \{0, h, 2h, \ldots\}$, and $\{Z_{nh}\}_{n\ge 0}$ is iid collection of standard normal random variables.
\item For realized stock price value $S_t$ at each $t$, we obtain the option prices with fixed strike price $S_t$  on a discrete grid using \eqref{eq11}. Finally to obtain the option price corresponding at $s=S_t$, we use linear interpolation of option prices at nearby grid points.

\item Using the numerical option price values, AIVP (as in Subsection 5.2) is obtained at each time point.
\item Finally, we plot both AIVP and the simulated $\sigma(X_t)$ against $t$ for comparing the dynamics.
\end{enumerate}
From the final plot we check if the AIVP time series recovers the transition times of hidden Markov chain.
\subsection{Experiments with Ideal Options having Fixed TTM}
We consider an MMGBM parameters with identical to those in Subsection \ref{numexp}, except the interest rate $r$ is set to be zero for simplicity of computation. Time step size $h=1/250$ and the initial asset price $S_0 =1$, and the Markov state $X_0 =1$ are taken for simulating the MMGBM till 200 time steps. We further set $p=1$ and $\tau =0.1$ for computing the AIVP for each time point.
\begin{figure}[!ht]
	\centering
	\includegraphics[width=0.6\textwidth]{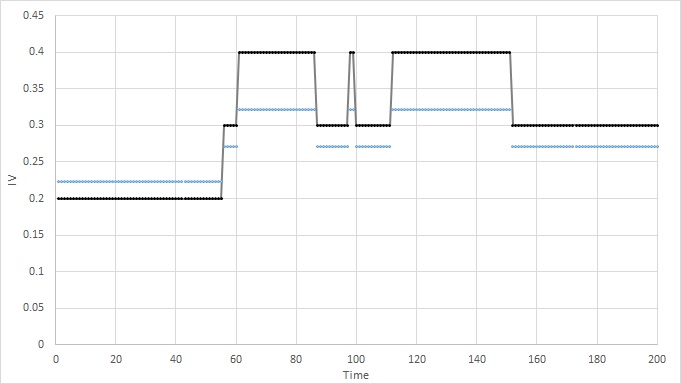}
	\caption{Comparison of AIVP and the simulated Markov chain}
	\centering
	\label{label:ATM IV time series}
\end{figure}
In Figure \ref{label:ATM IV time series} both AIVP and $\{\sigma(X_t)\}_t$ are plotted in blue and black respectively, where $X$ is the simulated Markov chain.
This plot shows that the jump discontinuities in the AIVP exactly match with those of the volatility coefficient of the underlying asset. In other words, for this numerical example, AIVP accurately recovers the instances of regime switching--not possible otherwise by analysing the asset process alone.
\subsection{Experiments with Realistic Options having Variable TTM}\label{IV4}
As not every day in a month is an expiration day of the call options, obtainability of IV time series with fixed TTM for all consecutive days is unrealistic. So, the approach in the preceding experiment has implementation limitations. In contrast, now we relax the condition of constant TTM and allow that to vary in an interval. One more implementation issue is the absence of perfect ATM options. Not all strike price contracts are available in the real market whereas stock price takes any values. In view of this, consideration of  $\{I_t^{S_t,p_tS_t,\tau_t}\}_{t\ge 0}$ instead of $\{I_t^{S_t,pS_t,\tau}\}_{t\ge 0}$ for defining AIVP, where $\{p_t\}_t$ and $\{\tau_t\}_t$ are allowed to vary in a narrow range, becomes prudent. In particular, if strike prices are in multiple of a constant $C$, then for example $p_t:=\rho(\frac{pS_t}{C}) \times \frac{C}{S_t}$ should be the choice for moneyness, where $\rho()$ is the \emph{round half down} function. Similarly, if the expiration times occur in multiples of $\beta$, then TTM could be $\tau_t:=\rho(\frac{t+\tau }{\beta}) \times\beta -t$. For the numerical experiment, we choose $p=1$, $\tau =0.12$, $C=0.01$, and $\beta = 0.08$. Since in days, $\tau$ and $\beta$ are 30 and 20 respectively, the TTM varies from 39 to 20 days. More precisely, if the first trading day has TTM 39 days, TTM decreases subsequently and becomes 20 on $20^{th}$ trading day. Once again twenty first day has contracts of TTM 39 days and so on.
\begin{figure}[h]
	\centering
	\includegraphics[width=0.6\linewidth]{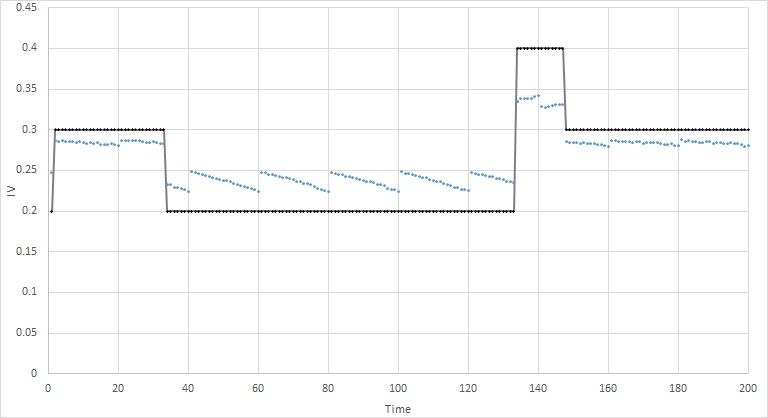}
	\caption{}
	\label{figiv4}
	\centering
\end{figure}
In Figure \ref{figiv4}, both the modified AIVP, i.e., $\{I_t^{S_t,p_tS_t,\tau_t}\}_{t\ge 0}$ and $\sigma(X_t)$ are plotted against $t$. While the former is shown in blue points, the latter is in black line. Vertical lines are drawn at every 20 days to indicate the expiry dates. A small shift in AIVP can be noticed right after the expiry dates, even though the market is in the same state.
Here market is in the same regime and small jump in AIVP appears due to a large change in TTM. Interestingly, AIVP is less than 0.25 or more than 0.3 if and only if $\sigma(X_t)$ is 0.2 or 0.4 respectively. Thus by identifying these cutoffs (in this case 0.25 and 0.3) the regimes may be recovered. In general such cutoffs can easily be obtained by a density based clustering. Besides, we can also notice that in state 1, where sigma value is 0.2, the IV decreases as the TTM decreases. But in state 3, which has sigma 0.4, the IV increases as the TTM decreases. In second state IV is almost constant throughout the expiry. These results are in accordance with observation illustrated in Figure \ref{IVTTM}.

In short, the above experiment indicates that the recovery of the regimes is possible even in the absence of option data across all strikes and all time to maturity.

\begin{figure}[h]
	\centering
	\includegraphics[width=0.6\linewidth]{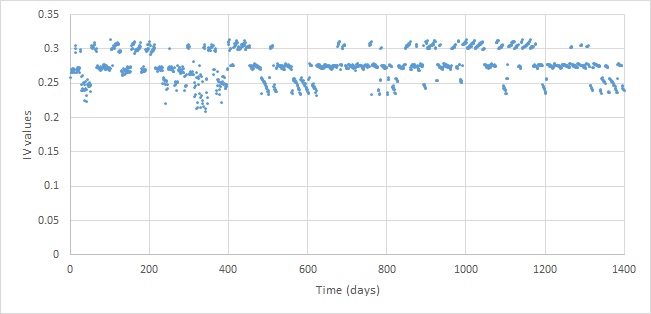}
	\caption{}
	\label{iv4plot}
	\centering
\end{figure}

\begin{figure}[h]
	\centering
	\includegraphics[width=0.6 \linewidth]{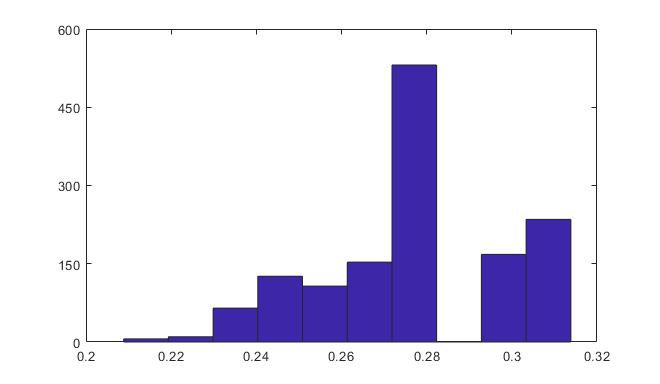}
	\caption{Histogram of the IV values. Number of bins is 10.}
	\label{hist}
	\centering
\end{figure}

\noindent \textbf{Quality of regime recovery:} Now we perform another numerical experiment with identical model parameters, where AIVP is obtained for 1400 consecutive time units (fig \ref{iv4plot}).
For recovering regimes we adopt the density based clustering. Two different local minima in the histogram of IV values are seen in Figure \ref{hist} of where the bin width is 0.01. By further refinement of binning, the local minima 0.26 and 0.28 are obtained and used as the cutoffs for different regimes. All occasions where the IV values are less than 0.26 or above 0.28 are assigned as Regime 1 or Regime 3 respectively. Regime 2 is assigned for the remaining occasions. The transitions identified by this method recovers actual transitions and correct regimes with $99\%$ accuracy.
\section{Conclusion}
\noindent It is important to note that the success of an usual breakpoint analysis of a time-series data  diminishes if the switching of the parameter values are not drastic or short lived, which are indeed the case for financial assets. Due to this, the applicability of the breakpoint analysis for recovering the volatility regimes from the asset price data, is somewhat limited. In view of this, the approach presented in this paper which gathers extra information from the derivative price data is advantageous. Theorem \ref{theo6} establishes a complete recovery of the abstract Market states from the option price data by considering the MMGBM setting. This theoretical result has also been validated using numerical examples. A novel element of the approach is the applicability of the proposed method even in the presence of several Market constraints beyond the theoretical settings, provided switching is not too fast and the volatility coefficient values at different regimes are not too close to each other. We have verified this by extensive numerical experiments. All codes are available at  https://github.com/agiiser/Regime-Recovery.

From the time series of recovered regimes, one can easily estimate the transition law of the hidden Markov chain. This gives a novel method of parameter estimation. Needless to mention, estimation of transition rate of a hidden Markov chain is a hard problem otherwise. Once, the transition parameters are estimated, the estimation of the volatility parameters for each regime becomes mathematically tractable. Further investigation is needed in this direction. Finding appropriate extension of the regime recovery is, in principle, another rich research agenda. We aim to extend this method for time in-homogeneous Markov and semi-Markov modulated GBM market models.

\noi {\bf{Acknowledgment} }
The authors are grateful to Milan Kumar Das for some very useful discussions.

\section{Appendix}\label{app}
\proof[Proof of Theorem \ref{ivtheorem}]
First we prove that for each fixed $t$, $s$, $K$ and $T$, $C( t,s;K,T;r,\cdot)$ is a bounded monotonic function and the range is an open interval given by $\big((s-K e^{-r(T-t)})^+,s\big)$. Next, we show that $M_p$ necessarily lies in the above range. We know that the function $C$ can be expressed using conditional expectation as below
\begin{equation}
      C( t,s;K,T;r,\sigma) = E\Big[e^{-r(T-t)}(\tilde{S}_T-K)^+\mid \tilde{S}_t=s\Big]
  \end{equation}
  where the process $\{\tilde{S}_t\}_{t\geq0}$ satisfies $d\tilde{S}_t=\tilde{S}_t(rdt+\sigma\*dW_t)$, $\tilde{S}_0>0$. Thus for the degenerate case, i.e, for
  $\sigma= 0$,
  $$C( t,s;K,T;r,0) = e^{-r(T-t)}(se^{-r(T-t)}-K)^+ = (s-Ke^{-r(T-t)})^+.$$
On the other hand, it is evident from \eqref{BSMF} that, as $\sigma \to \infty$, $d_1 \to \infty$ and $d_2 \to -\infty$. Hence,
$$\lim_{\sigma \to \infty}C( t,s;K,T;r,\sigma) =s.$$
Therefore, under strict monotonicity, $C$ is bounded with respect to $\sigma$, when other variables and parameters are fixed. To be more precise $C$ lies in $\big((s-K e^{-r(T-t)})^+,s\big)$. Furthermore using continuity of $C$ with respect to $\sigma$, for any fixed $t,s,K,T,r$, if the observed option value $M_p$ is in the range of $\big((s-Ke^{-r(T-t)})^+,s\big)$, there is a unique $0<\hat \sigma<\infty$ such that $C( t,s;K,T;r,\hat \sigma)=M_p$.

\noindent The continuity of $C$ w.r.t $\sigma$ follows from direct observation of BSM option pricing formula. The strict monotonicity follows from the positivity of the term $\frac{\partial C}{\partial \sigma}$, also known as Vega, which represents the change in the option price with respect to a unit change in volatility coefficient. From the BSM formula, a direct calculation by employing the partial derivative of the option price with respect to $\sigma$ shows that the Vega is always positive.  It is important to note that $M_p$, the fair value of the option in an efficient friction-less market never goes beyond the above mentioned interval under the no arbitrage (NA) assumption. We give a proof by contradiction method. To this end, first, assume
  $$0< M_p\leq (s-Ke^{-r(T-t)})^+,$$
  where the right side term is strictly positive under the above assumption. Hence the right side is the same as $(s-Ke^{-r(T-t)}).$ Therefore,
  $$0<Ke^{-r(T-t)}\leq s-M_p,$$
  or,
  \begin{equation}\label{IVproof}
    0<K\leq (s-M_p)e^{r(T-t)}.
  \end{equation}
Hence if a trader at time $t$ short sells one unit of stock to receive $s$ amount of money and purchases a call option at $M_p$ and keep the remaining amount in an ideal bank, s/he will get back $(s-M_p)e^{r(T-t)}$ amount of money at time $T$. Since this amount is not less than $K$ (see \eqref{IVproof}) s/he can, by exercising the option if needed, spend at most $K$ amount of money out of the bank's payoff to buy one unit of stock from the option writer or the stock market at time $T$. The stock s/he receives, can be returned to whom s/he borrowed from for the sake of short-selling. Thus the trader is still left with a non negative amount of money (which is positive with positive probability) with no chance of loss, leading to an arbitrage. On the other hand, if $M_p \geq s$, then a trader can write one unit of the call option to receive $M_p$ amount of money and buy one unit of stock at price $s$ and still be left with some non negative amount of money. At the maturity, if the option buyer exercises the option, the trader would give that one unit of stock to the buyer in exchange for $K$ amount of money. Thus, in the end, irrespective of any circumstances, the trader has certain non negative profit with no chance of loss, leading to an arbitrage. Therefore by the no arbitrage principle, we can say that the observed market price $M_p$ of the option lies in the above mentioned open interval. Hence we can compute IV by solving \eqref{IVdef} for $I$ for all the market traded options. \qed

\begin{lemma}\label{lemmaerror}
Assume $f(x)= \Big(1+\frac{a}{x}\Big)^x$ for $x>0$ and $a>0$.
Then
$$f(x)< e^a   \ \ \forall x>0.$$
\end{lemma}
\proof
Note that $f(1)=(1+a) < e^a$ and $\lim_{x \to \infty} f(x)= e^a$. Thus to prove the lemma it is sufficient to prove that $f^{\prime}(x)>0$ for all $x>0$. To this end, using $\ln f(x)= x\ln \left(1+\frac{a}{x}\right)$, we compute
\begin{align*}
\frac{d}{dx} \ln f(x)=\ln \Big(1+\frac{a}{x}\Big)+\frac{x}{\Big(1+\frac{a}{x}\Big)}\Big(\frac{-a}{x^2}\Big) = \ln \Big(1+\frac{a}{x}\Big)-\frac{\frac{a}{x}}{1+\frac{a}{x}} = \frac{1}{1+\frac{a}{x}}\Bigg[\Big(1+\frac{a}{x}\Big)\ln \Big(1+\frac{a}{x}\Big)-\frac{a}{x}\Bigg].
 \end{align*}
 Since the derivatives of $f$ and $\ln(f)$ have the same sign, we need to show
 $$\frac{d}{dx} \ln f(x) >0.$$
 The above expression implies that the derivative is positive if and only if the function $h$ given by
 $$h(y):=(1+y)\ln (1+y)-y \ \ \ \ $$
 is positive for all $y>0$. This is true because $h(0)=0$ and $$h^\prime (y) =\frac{1+y}{1+y}+\ln (1+y)-1=\ln (1+y)>0 \ \ \ \ \forall y>0.$$
 Hence the proof. \qed

\bibliography{IV.bbl}{}
 \bibliographystyle{siam}
\end{document}